\documentclass{elsart}
\usepackage{graphicx,graphics}
\usepackage{color}
\usepackage{amsmath}
\usepackage{amssymb}
\newcommand{\NP}{{\sf NP}}
\newcommand{\Pol}[0]{\ensuremath{\mathsf{Pol}}}

\newcommand{\reflex}[1]{{\mathcal #1}}

\newcommand{\SSS}{\mathcal{S}} 

\newcommand{\diam}{{\rm diam}}

\begin{document}

\begin{frontmatter}

\title{The Computational Complexity of Disconnected Cut and  $\mathbf{2K_2}$-Partition\thanksref{CP}}
\author{Barnaby Martin\thanksref{epsrc1} and Dani\"el Paulusma \thanksref{epsrc2}}

\address{
School of Engineering and Computing Sciences, Durham University,\\
  Science Labs, South Road, Durham DH1 3LE, U.K.}
\thanks[CP]{An extended abstract of this paper appeared in the proceedings of CP2011~\cite{MP11}.}
\thanks[epsrc1]{The first author was supported by EPSRC grant EP/G020604/1.}
\thanks[epsrc2]{The second author was supported by EPSRC grant EP/G043434/1.}

\maketitle

\begin{abstract}
For a connected graph $G=(V,E)$, a subset $U\subseteq V$ is called a disconnected cut if $U$
disconnects the graph and the subgraph induced by $U$ is disconnected as well. 
We show that the problem to test whether a graph  
has a disconnected cut is \NP-complete. 
This problem is polynomially equivalent to the following problems: testing if a graph has a $2K_2$-partition, testing if a graph allows a vertex-surjective homomorphism to the reflexive 4-cycle and testing if a graph has a spanning subgraph that consists of at most two bicliques. Hence, as an immediate consequence, these three decision problems are \NP-complete as well.
This settles an open problem frequently posed in each of the four settings.
\end{abstract}

\end{frontmatter}

\section{Introduction}

We solve an open problem that showed up as a missing case (often {\it the} missing case) in a  number of different research areas arising from connectivity theory, graph covers, graph homomorphisms
and graph modification. 
It is the only open question in the papers by Dantas et al.~\cite{DFGK05} and
Fleischner et al.~\cite{FMPS09},
a principal open question of Ito et al.~\cite{IKPT09,IKPT}, and the central question discussed
by  Cook et al.~\cite{CDEFFK10} and  Dantas et al.~\cite{DMS10}.
Indeed, the problem is considered important enough to generate its own complexity class~\cite{F,TDF10}, and it is known to be tractable for many graph classes~\cite{CDEFFK10,DMS10, FMPS09,IKPT09}.

Before we explain how these areas are related, we briefly describe them first.
Throughout the paper, 
we consider undirected finite graphs that have  no multiple edges.
Unless explicitly stated otherwise they do not have self-loops either.
We denote the vertex set and edge set of a graph $G$ by $V_G$ and $E_G$, respectively.
If no confusion is possible, we may omit the subscripts. 
We let $n=|V(G)|$ denote the number of vertices of $G$. 
The {\it complement} of a graph $G=(V,E)$ 
is the graph $\overline{G}=(V,\{uv\notin E\; |\; u\neq v\})$.
For a subset $U \subseteq V_G$, we let $G[U]$ denote the subgraph of $G$ {\it induced by} $U$, which is the graph $(U,\{uv\;  |\; u,v\in U\; \mbox{and}\; uv\in E_G\}$).

\subsection{Vertex Cut Sets}

A maximal connected subgraph of $G$ is called a {\em component} of $G$. 
A {\it vertex cut (set)} or {\it separator} of a graph $G=(V,E)$ is a subset $U\subset V$ 
 such that $G[V\backslash U]$ contains at least two components.
 
Vertex cuts play an important role in graph connectivity, and various kinds of 
vertex cuts have been studied in the literature. 
For instance, a cut $U$ of a graph $G=(V,E)$ is called a {\it $k$-clique cut} 
if $G[U]$ has a spanning subgraph consisting of $k$ complete graphs;
a {\it strict $k$-clique cut} if $G[U]$ consists of $k$ components that are complete graphs;
a {\it stable cut} if $U$ is an independent set; and 
a {\it matching cut} if $E_{G[U]}$ is a matching. 
The problem that asks whether a graph has a $k$-clique cut is solvable in polynomial time for $k=1$ and $k=2$, as shown by
Whitesides~\cite{Wh81} and Cameron et al.~\cite{CEHS07}, respectively.
The latter authors also showed that deciding if a graph has a strict 2-clique cut can be solved in polynomial time.
On the other hand,
the problems that ask whether a graph has a 
stable cut or a matching cut, respectively, 
are \NP-complete, as shown by Chv\'atal~\cite{Ch84} and Brandst\"adt et al.~\cite{BDBS00}, respectively.

For a fixed constant $k \ge 1$, a cut $U$ of a connected graph $G$ is called a {\em $k$-cut} of $G$ if $G[U]$ contains exactly $k$ components. 
Testing if a graph has a $k$-cut is solvable in polynomial time for $k=1$, whereas it is \NP-complete for every fixed $k\geq 2$~\cite{IKPT09}.
For $k\geq 1$ and $\ell\geq 2$, a $k$-cut $U$ is called a $(k,\ell)$-{\it cut} of a graph $G$ if $G[V\backslash U]$ consists of exactly $\ell$ components.
Testing if a graph has a $(k,\ell)$-cut is polynomial-time solvable when $k=1$, $\ell \ge 2$,
and  \NP-complete otherwise~\cite{IKPT09}. 

A cut $U$ of a graph $G$ is called \emph{disconnected} if $G[U]$ contains at least two components.
We observe that $U$ is a disconnected cut if and only if 
$V\backslash U$ is a disconnected cut if and only if $U$ is a $(k,\ell)$-cut for some $k\geq 2$ and $\ell\geq  2$. The following question was posed in several papers~\cite{FMPS09,IKPT09,IKPT} 
as an open problem.

\medskip
\noindent
{\it Q1. How hard is it to test if a graph has a disconnected cut?}

\medskip
\noindent
The problem of testing if a graph has a disconnected cut 
is called the {\sc Disconnected Cut} problem. 
It is known that every graph of diameter 1 has no disconnected cut, and every graph of diameter at least 3 has a disconnected cut~\cite{FMPS09}. Hence, in order to determine the computational complexity of {\sc Disconnected Cut}, we may restrict ourselves to graphs of diameter 2. 

A disconnected cut $U$ of a connected graph $G=(V,E)$ is {\it minimal}
if $G[(V\backslash U) \cup \{u\}]$ is connected for every $u \in U$. Recently, the corresponding decision problem called {\sc Minimal Disconnected Cut} was shown to be \NP-complete~\cite{IKPT}.

\subsection{$H$-partitions}

A \emph{model graph} $H$ with $V_H=\{h_0,\ldots,h_{k-1}\}$ has two types of edges: solid and dotted edges, and an {\it $H$-partition} of a graph $G$ is a
partition of $V_G$ into $k$ (nonempty) sets $V_0,\dots,V_{k-1}$ such that
for all vertices $u\in V_i$, $v\in V_j$ and for all $0\leq i<j\leq k-1$ the following two conditions hold.
Firstly, if $h_ih_j$ is a solid edge of $H$, then $uv\in E_G$.
Secondly, if $h_ih_j$ is a dotted edge of $H$, then $uv\notin E_G$.
There are no such restrictions when $h_i$ and $h_j$ are not adjacent. 
Let $2K_2$ be the model graph with vertices
$h_0,\dots,h_3$, solid edges $h_0h_2, h_1h_3$ and no dotted edges, and
$2S_2$ be the model graph with vertices
$h_0,\dots,h_3$, dotted edges $h_0h_2, h_1h_3$ and no solid edges.
We observe that a graph $G$ has a $2K_2$-partition if and only if 
its complement $\overline{G}$ has a $2S_2$-partition.

The following question was mentioned in several papers~\cite{CDEFFK10,DFGK05,DMS10,F,TDF10} as an open problem.

\medskip
\noindent
{\it Q2. How hard is it to test if a graph has a $2K_2$-partition?}

\medskip
\noindent
One of the reasons for posing this question is that the (equivalent) cases $H=2K_2$ and $H=2S_2$ are the only two cases of model graphs on at most four vertices for which the computational complexity of the corresponding decision problem, called $H$-{\sc Partition}, is still open.
In fact it is known that $H$-{\sc Partition} is polynomial-time solvable for all other 4-vertex model graphs~$H$.
In particular, the model graph $H$ with vertices $h_0,\ldots,h_3$, solid edge $h_0h_2$ and dotted edge $h_1h_3$ is well known. In that case the $H$-{\sc Partition} problem is called the {\sc Skew Partition} problem.
Note that this problem is equivalent to asking whether the vertex set of a given graph can be partitioned into
two sets $V_1$ and $V_2$ such that $V_1$ induces a disconnected graph in~$G$ and $V_2$ induces
a disconnected graph in~$\overline{G}$.
Even the list version of this problem, where each vertex has been assigned a list of blocks in which it must be placed, is polynomial-time solvable, as shown by de Figueiredo, Klein and Reed~\cite{FKR00} (later, Kennedy and
Reed~\cite{KR08} presented a faster polynomial-time algorithm for the non-list version).
All other cases of $H$-{\sc Partition} for 4-vertex model graphs~$H\notin \{K_2,S_2\}$ have been settled by Dantas et al.~\cite{DFGK05}. 

In the literature, $2K_2$-partitions have been well studied, see e.g. 
three  recent papers of
Cook et al.~\cite{CDEFFK10},
Dantas, Maffray and Silva~\cite{DMS10} and Teixeira, Dantas and de Figueiredo~\cite{TDF10}. The first two papers~\cite{CDEFFK10,DMS10} study the $2K_2$-{\sc Partition} problem for several graph classes, and the third paper~\cite{TDF10} defines a new class of problems called $2K_2$-hard. 
In addition, the first paper also proves that $2K_2$-{\sc Partition} can be solved in $O((2^d-1)n^2)$ time for $n$-vertex graphs of minimum vertex degree~$d$.
By a result on retractions of Hell and Feder~\cite{FH98}, which we explain later, 
the list versions of $2S_2$-{\sc Partition} and $2K_2$-{\sc Partition}  
are \NP-complete.
A variant on $H$-partitions that allows empty blocks $V_i$ in an $H$-partition is studied by Feder et al.~\cite{FHKM03}, whereas Cameron et al.~\cite{CEHS07} consider the list version of this variant.

\subsection{Graph Covers}

Let $G$ be a graph and  $\SSS$ be a set of (not
necessarily vertex-induced) subgraphs of $G$ that has size $|{\cal S}|$. 
The set $\SSS$ is a
\emph{cover} of $G$ if every edge of $G$ is
contained in at least one of the subgraphs in $\SSS$.  The set $\SSS$
is a \emph{vertex-cover} of $G$ if every vertex of $G$ is contained in
at least one of the subgraphs in $\SSS$.  If all subgraphs in $\SSS$
are \emph{bicliques}, that is, complete connected bipartite graphs,
then we speak of a \emph{biclique cover} or a \emph{biclique
vertex-cover}, respectively.  
Testing whether a graph has a biclique cover of size at most $k$ is polynomial-time solvable for 
any fixed $k$; it is even fixed-parameter tractable in $k$ as shown by Fleischner et al.~\cite{FMPS09}. The same authors~\cite{FMPS09} show that 
testing whether a graph has a biclique vertex-cover of size at most $k$ is polynomial-time solvable for $k=1$ and \NP-complete for $k\geq 3$. For $k=2$, they show that this problem can be solved in polynomial time for bipartite input graphs, and they pose the following open problem.

\medskip
\noindent
{\it Q3. How hard is it to test if a graph has a biclique vertex-cover of size
$2$?}

\medskip
\noindent
The problem of testing if a graph has a biclique vertex-cover of size 2 is called the 
2-{\sc Biclique Vertex-Cover} problem.
In order to answer question Q3 we may without loss of generality restrict to biclique vertex-covers in which
every vertex is in exactly one of the subgraphs in $\SSS$ (cf.~\cite{FMPS09}).

\subsection{Graph Homomorphisms}

A {\it  homomorphism} from a graph $G$ 
to a graph $H$ is a mapping $f: V_G \to V_H$ that maps adjacent vertices of $G$ to adjacent vertices of $H$, 
i.e., $f(u)f(v)\in E_H$ whenever $uv\in E_G$.
The problem {\sc $H$-Homomorphism} tests whether a given graph $G$ 
(also called the {\it guest graph})
allows a homomorphism to a 
graph $H$ called the {\it target} which is fixed, i.e., not part of the input.  This problem is also
known as $H$-{\sc Coloring}.
Hell and Ne\v{s}et\v{r}il~\cite{HN90} showed that $H$-{\sc Homomorphism} is solvable in polynomial time if $H$ is bipartite, and \NP-complete otherwise.
Here, $H$ is assumed not to have a self-loop $xx$, 
as otherwise we can map every vertex of $G$ to $x$.

A homomorphism $f$ from a graph $G$ to a graph $H$ is {\it surjective} if for each $x\in V_H$ there exists at least one vertex $u\in V_G$ with $f(u)=x$.
This leads to the problem of deciding if a given graph allows a surjective homomorphism to a fixed target graph $H$, which is called the {\sc Surjective $H$-Homomorphism} or 
{\sc Surjective $H$-Coloring} problem.
For this variant, the presence of a vertex with a self-loop in the target graph $H$ does not make the problem trivial. Such vertices are
called {\it reflexive}, whereas vertices with no self-loop are said to be {\it irreflexive}. 
A graph  that contains 
one or more reflexive vertices is called {\it partially reflexive}.
In particular, a graph is {\it reflexive} if all its vertices are reflexive, and a graph is {\it irreflexive} if all its vertices are irreflexive.
Golovach, Paulusma and Song~\cite{GPS11} showed  that for any fixed tree $H$, the
{\sc Surjective $H$-Homomorphism} problem is polynomial-time solvable if 
the (possibly empty) set of reflexive vertices in $H$ induces a connected subgraph of $H$, and \NP-complete otherwise. They mention that 
the smallest open case is  the case, in which $H$ is the reflexive 
4-cycle denoted ${\cal C}_4$. 

\medskip
\noindent
{\it Q4. How hard is it to test if a graph has a surjective homomorphism to ${\cal C}_4$?}

\medskip
\noindent
The following two notions are closely related to surjective homomorphisms.
A homomorphism $f$ from a graph $G$ to an induced subgraph $H$ of $G$
is a {\it retraction} from $G$ to $H$ if $f(h)=h$ for all $h\in V_H$. 
In that case we say that $G$ {\it retracts to} $H$.
Note that this implies that $G$ allows a vertex-surjective homomorphism 
to $H$, whereas the reverse implication does not necessarily hold.
For a fixed graph $H$, the $H$-{\sc Retraction} problem has as input a graph $G$ that contains $H$ as an induced subgraph and is to test if $G$ retracts to $H$. 
Hell and Feder~\cite{FH98} showed that $\reflex{C}_4$-{\sc Retraction}
is \NP-complete. 

We emphasize that a surjective homomorphism is  vertex-surjective. A stronger notion is to require a homomorphism from a graph $G$ to a graph $H$ to be {\it edge-surjective}, which means that for any edge $xy\in E_H$ with $x\neq y$ there exists an edge $uv\in E_G$ with $f(u)=x$ and $f(v)=y$. Note that the edge-surjectivity
condition only holds for edges $xy\in E_H$; there is no such condition on the self-loops $xx\in E_H$.
An edge-surjective homomorphism is also called a {\it compaction}.
If $f$ is a compaction from $G$ to $H$, we say that $G$ {\it compacts} to $H$.
The $H$-{\sc Compaction} problem asks if a graph $G$ compacts to a 
fixed graph~$H$.
Vikas~\cite{Vi02,Vi04,Vi05} determined the computational complexity of this problem for several classes of fixed target graphs. In particular, he showed that $\reflex{C}_4$-{\sc Compaction} is \NP-complete~\cite{Vi02}.
More recently, Vikas~\cite{Vi11,Vi13}  considered $H$-{\sc Compaction} for
guest graphs belonging to some restricted graph class.

\subsection{Graph Contractibility}\label{s-contract}

A graph modification problem has as input a graph $G$ and an integer $k$.
The question is whether $G$ can be modified to belong to some specified graph class
that satisfies further properties by using at most $k$ operations of a certain specified type such as deleting a vertex or deleting an edge.
Another natural operation is the {\it contraction} of an edge, 
which removes both end-vertices of the edge and replaces them by a
new vertex adjacent to precisely those vertices 
that were adjacent to at least one of the two end-vertices.
If a graph $H$ can be obtained from $G$ by a sequence of edge contractions, 
then $G$ is said to be {\it contractible to} $H$. 
The problem {\sc $\Pi$-Contractibility} has as input a graph $G$ together with an integer $k$ and is to test whether $G$ is contractible to a graph in $\Pi$ by using at most $k$ edge contractions. 

Asano and Hirata~\cite{AH83} show that  $\Pi$-{\sc Contractibility} is \NP-complete if $\Pi$ satisfies certain
conditions. As a consequence,  this problem is \NP-complete for many graph classes $\Pi$ such as the classes of planar graphs, outerplanar graphs, series-parallel graphs, forests and chordal graphs.
By a result of 
Heggernes et al.~\cite{HHLLP11},  $\Pi$-{\sc Contractibility} is \NP-complete for trees even if  the input graph is bipartite.
 If $\Pi$ is the class of paths or cycles, then $\Pi$-{\sc Contractibility}
is polynomially equivalent to the problems of determining the length of a longest path and a longest cycle, respectively, to which a given graph can be contracted. The first problem has been shown to be \NP-complete by van 't Hof,  Paulusma and Woeginger~\cite{HPW09} even for graphs with no induced path on 6 vertices. The second problem has been shown to be \NP-complete by Hammack~\cite{Ha02}. Heggernes et al.~\cite{HHLP11} observed that $\Pi$-{\sc Contractibility} is \NP-complete when $\Pi$ is the class of bipartite graphs, whereas
Golovach et al.~\cite{GKPT11} showed that $\Pi$-{\sc Contractibility} is \NP-complete
when $\Pi$ is the class of graphs of a certain minimum degree $d$; they show that $d=14$ suffices.

A graph $G$ contains a graph $H$ as a {\it minor} if $G$ can be modified to $H$ by a sequence of 
vertex deletions, edge deletions, and edge contractions.  
Eppstein~\cite{Ep09} showed that it is \NP-complete to decide if a given graph $G$ has a complete graph $K_h$ as a minor for some given integer $h$. 
This problem is equivalent to deciding if a graph $G$ is contractible to $K_h$. 
Hence, $\Pi$-{\sc Contractibility} is \NP-complete if $\Pi$ is the class of complete graphs.

The biclique with partition classes of size $k$ and $\ell$, respectively,  is denoted $K_{k,\ell}$.
A {\it star} is a biclique $K_{1,\ell}$ for some integer $\ell\geq 2$. 
If $\Pi$ is the class of stars, then $\Pi$-{\sc Contractibility} is \NP-complete. This can be seen as follows. 
Let $K_1\Join G$ denote the graph obtained from a graph $G$ after adding a new vertex and making it adjacent to all vertices of $G$.
Then $G$ has an independent set of size $h$ for some given integer $h\geq 2$ if and only if $K_1\Join G$ is contractible to a star $K_{1,h}$. Since the first problem is \NP-complete~\cite{GJ79,Ka72}, the result follows.

A remaining elementary graph class is the class of bicliques $K_{k,\ell}$ with
$k\geq 2$ and $\ell\geq 2$; we call such bicliques {\it proper}.
In order to determine the complexity for this graph class, we first consider the following question. 

\medskip
\noindent
{\it Q5. How hard is it to test if a graph is contractible to a  proper biclique?}

The problem of testing whether a graph can be contracted to a  proper biclique is called the {\sc Biclique Contractibility} problem. By setting $k=n$, we see that this problem is a special instance of the corresponding $\Pi$-{\sc Contractibility} problem. If one of the two integers $k\geq 2$ or $\ell\geq 2$ is fixed,  then testing if $G$ is contractible to $K_{k,\ell}$ is known to be \NP-complete~\cite{IKPT09}.

\subsection{The Relationships Between Questions Q1--Q5}

Before we explain how questions Q1--Q5 are related, we 
introduce some new terminology.
The {\em distance} $d_G(u,v)$ between two vertices $u$ and $v$ in a graph $G$
is the number of edges in a shortest path between them;
if there is no path between $u$ and $v$ then $d_G(u,v)=\infty$.
The {\em diameter} $\diam(G)$ is defined as $\max\{d_G(u,v)\; |\; u,v\in V\}$.
A biclique is called {\it nontrivial} if $k\geq 1$ and $\ell\geq 1$.

\begin{prop}[\cite{IKPT09}]\label{p-known}
Let $G$ be a connected graph.
Then statements $(1)$--$(5)$ are equivalent:
\begin{itemize}
\item [$(1)$] $G$ has a disconnected cut.
\item [$(2)$] $G$ has a {\it $2S_2$-partition}.
\item [$(3)$] $G$ allows a vertex-surjective homomorphism to 
$\reflex{C}_4$.
\item [$(4)$]
$\overline{G}$ has a spanning subgraph
  that consists of exactly two nontrivial bicliques.
\item [$(5)$] $\overline{G}$ has a {\it $2K_2$-partition}.
\end{itemize}
If $\diam(G)=2$, then $(1)$--$(5)$ are also equivalent to the following statements: 
\begin{itemize}
\item [$(6)$]
$G$ allows a compaction to $\reflex{C}_4$.
\item [$(7)$]
$G$ is contractible to some biclique $K_{k,\ell}$ for some $k,\ell\geq 2$. 
\end{itemize}
\end{prop}

Due to Proposition~\ref{p-known}, questions Q1--Q4 are equivalent.
Recall that we may restrict ourselves to graphs of diameter~2, as every graph of diameter 1 has no disconnected cut, and every graph of diameter at least 3 has a disconnected cut~\cite{FMPS09}. Under this restriction, Proposition~\ref{p-known} tells us that Q1--Q4 are also equivalent to~Q5 and
to the question of determining the computational complexity of $\reflex{C}_4$-{\sc Compaction}.
Recall that Vikas~\cite{Vi02} showed that the latter problem is \NP-complete. However,  the gadget in his \NP-completeness reduction has diameter~3 as observed by Ito et al.~\cite{IKPT}.
 
\medskip
\noindent
{\bf Our Result.} 
 A pair of vertices in a graph is a {\it dominating (non-)edge} if the two vertices of the pair are 
(non-)adjacent, and any other vertex in the graph is adjacent to at least one of them. 
We solve question Q4 by showing that the problem {\sc Surjective ${\cal C}_4$-Homomorphism} is indeed 
\NP-complete
for graphs of diameter 2 even if they have a dominating non-edge.
 In contrast, Fleischner et al.~\cite{FMPS09} showed that this problem is polynomial-time solvable on input graphs with a dominating edge.
As a consequence of our result 
and Proposition~\ref{p-known}, 
we find that the problems {\sc Disconnected Cut}, $2K_2$-{\sc Partition}, $2S_2$-{\sc Partition}, and  2-{\sc Biclique Vertex-Cover} and 
also that
the problems ${\cal C}_4$-{\sc Compaction} and {\sc Biclique Contraction} are \NP-complete for graphs of diameter 2
even if they have a dominating non-edge.
Hence, we have not only solved question~Q4 but also questions~Q1,~Q2,~Q3 and~Q5.

Our approach to prove \NP-completeness is as follows. As mentioned before, we can restrict ourselves to graphs of diameter 2. We therefore try to reduce the diameter in the gadget of the \NP-completeness proof of Vikas~\cite{Vi02} for $\reflex{C}_4$-{\sc Compaction} from 3 to 2. This leads to \NP-completeness of {\sc Surjective $\reflex{C}_4$-Homomorphism}, because these two problems coincide for graphs of diameter 2 due to Proposition~\ref{p-known}. 
The proof that $\reflex{C}_4$-{\sc Compaction}
is \NP-complete~\cite{Vi02} has its roots in the proof that $\reflex{C}_4$-{\sc Retraction} is \NP-complete~\cite{FH98}. So far, it was only known that
$\reflex{C}_4$-{\sc Retraction} stays \NP-complete for graphs of diameter 3~\cite{IKPT}.
We start our proof by showing that $\reflex{C}_4$-{\sc Retraction} is \NP-complete even for graphs of diameter 2. The key idea is to base the reduction from an \NP-complete homomorphism (constraint satisfaction) problem that we obtain only after a fine analysis under the algebraic conditions of 
Bulatov~\cite{Conservative} and
Bulatov, Krokhin and Jeavons \cite{JBK}, which we perform in Section~\ref{s-algebra}.
This approach is novel in the sense that usually 
graph theory provides a test-bed for constraint satisfaction problems whereas here we see a case where the flow of techniques is the other way around.
We present our \NP-completeness proof for $\reflex{C}_4$-{\sc Retraction} on graphs of diameter 2 in Section~\ref{s-retract}. 
This leads to a special input graph of the $\reflex{C}_4$-{\sc Retraction} problem, which enables us to modify the gadget of the proof of Vikas~\cite{Vi02} for $\reflex{C}_4$-{\sc Compaction} in order to get its diameter down to 2, as desired. We explain this part in Section~\ref{s-surjective}.

We also point out that Vikas~\cite{Vi11,Vi13} has announced to have an \NP-completeness proof of 
{\sc $\reflex{C}_4$-Homomorphism} as well but so far has not made his proof  publicly available.

\section{Constraint Satisfaction}\label{s-algebra}
 
The notion of a graph homomorphism can be generalized as follows.
A {\it structure} is a tuple ${\cal A}=(A; R_1,\ldots, R_k)$, 
where $A$ is a set called the {\it domain} of ${\cal A}$
and $R_i$ is an $n_i$-ary {\it relation} on $A$ for $i=1,\ldots, k$, i.e.,
a set of $n_i$-tuples of elements from $A$.
Note that a graph $G=(V,E)$ can be seen as a structure $G=(V;\{(u,v),(v,u)\; |\; uv\in E\})$. 
Throughout the paper we only consider {\it finite} structures, i.e., with a finite domain.

Let ${\cal A}=(A; R_1,\ldots, R_k)$ and ${\cal B}=(B; S_1,\ldots,S_k)$ be two structures, 
where each $R_i$ and $S_i$ are relations of the same arity $n_i$. 
Then a {\it homomorphism} from ${\cal A}$ to ${\cal B}$ is a mapping
$f: A\rightarrow B$ such that $(a_1 , \ldots, a_{n_i})
\in R_i$ implies $(f(a_1 ), \ldots, f (a_{n_i})) \in S_i$
for every $i$ and every $n_i$-tuple
$(a_1, \ldots, a_{n_i}) \in A^{n_i}$.
The decision problem that is to test if a given structure ${\cal A}$ allows a homomorphism to 
a fixed structure ${\cal B}$ is called the
${\cal B}$-{\sc Homomorphism} problem,
also known as the  ${\cal B}$-{\sc Constraint Satisfaction} problem.

Let ${\cal A}=(A; R_1,\ldots, R_k)$ be a structure
and $\ell$ be an integer.
The {\it power structure} $\mathcal{A}^\ell$ has domain $A^\ell$ and
for $1\leq i\leq k$, has relations 
$$R^\ell_i:=\{((a^1_1,\ldots,a^1_\ell),\ldots,(a^{n_i}_1,\ldots,a^{n_i}_\ell))\; |\;
(a^1_1,\ldots,a^{n_i}_1),\ldots,  (a^1_\ell,\ldots,a^{n_i}_\ell) \in R_i\}.$$
We note that $R^1_i=R_i$ for $1\leq i\leq k$.
An ($\ell$-ary) \emph{polymorphism} of  $\mathcal{A}$ is a homomorphism from $\mathcal{A}^\ell$ to $\mathcal{A}$ for some integer $\ell$. A $1$-ary polymorphism is also called an {\it endomorphism}. The set of polymorphisms of $\mathcal{A}$ is denoted Pol$(\mathcal{A})$.
 
A binary function $f$ on a domain $A$ is a {\it semilattice} function if $f(h,f(i,j))$ $= $ $f(f(h,i),j)$, $f(i,j)=f(j,i)$, and $f(i,i)=i$ for all $i,j\in A$. 
A ternary function $f$ is a {\it Mal'tsev} function if
$f(i,j,j)=f(j,j,i)=i$ for all $i,j\in A$.
A ternary function $f$ is a \emph{majority} function if
$f(h,h,i)=f(h,i,h)=f(i,h,h)=h$ for all $h,i\in A$.
On the Boolean domain $\{0,1\}$, we may consider propositional functions.
The only two semilattice functions on the Boolean domain are
the binary function $\wedge$, which maps $(h,i)$ to $(h \wedge i)$, 
which is $1$ if $h=i=1$ and $0$ otherwise,
and the binary function $\vee$ which maps $(h,i)$ to $(h \vee i)$,
which is $0$ if $h=i=0$ and $1$ otherwise. 
We may consider each of these two functions on any two-element domain (where we view one element as $0$ and the other as $1$). 
For a function $f$ on $B$, and a subset $A \subseteq B$, we let $f_{|A}$ be the restriction of $f$ to $A$.

A structure is a \emph{core} if all of its endomorphisms are {\it automorphisms}, i.e., are invertible. We will make use of the following theorem from Bulatov, Krokhin and Jeavons~\cite{JBK} (it appears in this form in 
Bulatov~\cite{Conservative}).

\begin{thm}[\cite{Conservative,JBK}]
\label{thm:BJK}
Let $\mathcal{B}=(B;S_1,\ldots,S_k)$ be a core and $A \subseteq B$ be a subset of size $|A|= 2$ that as a unary relation is in $\mathcal{B}$. If for each $f \in \Pol(\mathcal{B})$, $f_{|A}$ is not majority, semilattice or Mal'tsev, then ${\cal B}$-{\sc Homomorphism} is \NP-complete.
\end{thm}

Let $\mathcal{D}$ be the structure on domain $D=\{0,1,3\}$ with four binary relations
\[\begin{array}{lcl}
S_1 &:= &\{(0,3),(1,1),(3,1),(3,3)\}\\[3pt] 
S_2 &:= &\{(1,0),(1,1),(3,1),(3,3)\}\\[3pt]
S_3 &:= &\{(1,3),(3,1),(3,3)\}\\[3pt]
S_4 &:=&\{(1,1),(1,3),(3,1)\}.
\end{array}\]
We use $\{0,1,3\}$ (instead of, say, $\{0,1,2\}$) to tie in exactly with the Vikas~\cite{Vi02} labelling of $\reflex{C}_4$.
\begin{prop}\label{p-d4}
The ${\mathcal D}$-{\sc Homomorphism} problem is \NP-complete.
\end{prop}

\begin{pf}
We use Theorem~\ref{thm:BJK}.
We first show that ${\cal D}$ is a core.
Let $g$ be an endomorphism of ${\cal D}$. 
We must show that $g$ is an automorphism.
If $g(0)=3$ then $g(1)=3$ by preservation of $S_2$, i.e., as otherwise $(1,0)\in S_2$ does not
imply $(g(1),g(0))\in S_2$. However, $(1,1)\in S_4$ but 
$(g(1),g(1))=(3,3)\notin S_4$.
Hence $g(0)\neq 3$.
If $g(0)=1$ then $g(3)=1$ by preservation of $S_1$.
However, $(3,3)\in S_3$ but $(g(3),g(3))=(1,1)\notin S_3$.
Hence $g(0)\neq 1$. This means that
$g(0)=0$. Consequently, $g(1)=1$ by preservation of $S_2$, and $g(3)=3$
by preservation of $S_1$. 
Hence, $g$ is the identity mapping, which is an automorphism, as desired.

Let $A=\{1,3\}$, which is in $\mathcal{D}$ in the form of $S_1(p,p)$ (or $S_2(p,p)$). Suppose that $f\in \Pol(\mathcal{D})$.
In order to prove Proposition~\ref{p-d4}, we must show that
$f_{|A}$ is neither majority nor semilattice nor Mal'tsev.

Suppose that $f_{|A}$ is semilattice. Then $f_{|A}=\wedge$ or
$f_{|A}=\vee$.
If $f_{|A}=\wedge$, then 
either $f(1,1)=1$, $f(1,3)=3$, $f(3,1)=3$, $f(3,3)=3$, or
$f(1,1)=1$, $f(1,3)=1$, $f(3,1)=1$, $f(3,3)=3$ depending on how
the elements $1,3$ correspond to the two elements of the Boolean domain.
The same holds if $f_{|A}=\vee$.
Suppose that $f(1,1)=1$, $f(1,3)=3$, $f(3,1)=3$, $f(3,3)=3$.
By preservation of $S_4$ we find that $f(1,3)=1$ due to $f(3,1)=3$.
This is not possible, as $f(1,3)=3$.
Suppose that $f(1,1)=1$, $f(1,3)=1$, $f(3,1)=1$, $f(3,3)=3$. 
By preservation of $S_3$ we find that $f(1,3)=3$ due to $f(3,1)=1$.
This is not possible either.

Suppose that $f_{|A}$ is Mal'tsev.  
By preservation of $S_4$, we find that $f(1,1,3)=1$ due to $f(3,1,1)=3$.
However, because $f(1,1,3)=3$, this is not possible.

Suppose that $f_{|A}$ is majority.
By preservation of $S_1$, we deduce that $f(0,3,1)\in \{0,3\}$ due to
$f(3,3,1)=3$, and that $f(0,3,1)\in \{1,3\}$ due to $f(3,1,1)=1$. 
Thus, $f(0,3,1)=3$. By preservation of $S_2$, however, we deduce
that $f(0,3,1)\in \{0,1\}$ due to $f(1,3,1)=1$.
This is a contradiction. Hence, we have completed the proof of Proposition~\ref{p-d4}.\qed
\end{pf}

\section{Retractions}\label{s-retract}

In the remainder of this paper, the graph $H$ denotes the reflexive 4-vertex
cycle $\reflex{C}_4$. 
We let $h_0,\ldots,h_3$ be the vertices and $h_0h_1$, $h_1h_2$, $h_2h_3$, and $h_3h_0$ be
the edges of $H$. We prove that $H$-{\sc Retraction} is \NP-complete for graphs of diameter 2 by a reduction from ${\cal D}$-{\sc Homomorphism}.

Let ${\cal A}=(A;R_1,\ldots,R_4)$ be an instance of ${\cal D}$-{\sc Homomorphism}, where we may assume that each $R_i$ is a binary relation.
From ${\cal A}$ we construct a graph $G$ as follows.
We let the elements in ${\cal A}$ correspond to vertices of $G$. 
If $(p,q)\in R_i$ for some $1\leq i\leq 4$, then we say that vertex $p$ in $G$ is of {\it type} $\ell$ and vertex $q$ in $G$ is of {\it type} $r$.
Note that a vertex can be of type $\ell$ and $r$ simultaneously, because it can be the first element in a pair in $R_1\cup \cdots \cup R_4$ and the second element of another such pair.
For each $(p,q)\in R_i$ and $1\leq i\leq 4$ we introduce four new vertices
$a_p,b_p,c_q,d_q$ with edges $a_pp$, $a_pb_p$, $b_pp$, $c_qq$, $c_qd_q$ and $d_qq$. We say that a vertex $a_p,b_p,c_q,d_q$ is of {\it type} $a,b,c,d$, respectively; note that these vertices all have a unique type.

We now let the graph $H$ be an induced subgraph of $G$ (with distinct vertices $h_0,\ldots,h_3$). 
Then formally $G$ must have self-loops $h_0h_0,\ldots, h_3h_3$, 
because $H$ has such self-loops.
Outside of $H$ in $G$, it does not matter whether we consider~$G$ to have self-loops or not. In any case we do not draw any loops in our figures in order to keep these uncluttered. 

In $G$ we join every $a$-type vertex to $h_0$ and $h_3$, every $b$-type vertex
to $h_1$ and $h_2$, every $c$-type vertex to $h_2$ and $h_3$, and every
$d$-type vertex to $h_0$ and $h_1$. We also add an edge between $h_0$ and every vertex of $A$.

\begin{figure}
\begin{minipage}[b]{0.5\linewidth}
\begin{center}
\includegraphics[scale=0.9]{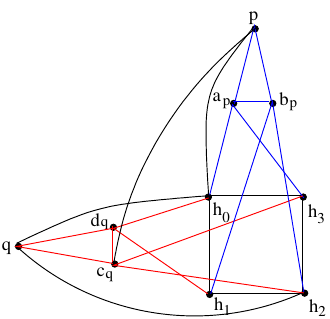}
\caption{The part of a ${\cal D}$-graph $G$ for a pair $(p,q)\in R_1$.}\label{f-I}
\end{center}
\end{minipage}
\hspace{0.5cm}
\begin{minipage}[b]{0.5\linewidth}
\begin{center}
\includegraphics[scale=0.9]{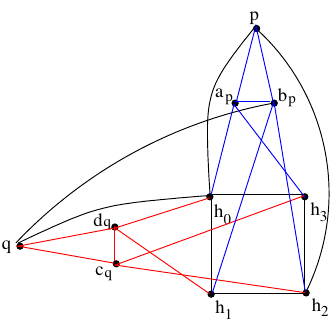}
\caption{The part of a ${\cal D}$-graph $G$ for a pair $(p,q)\in R_2$.}\label{f-II}
\end{center}
\end{minipage}
\end{figure}

\begin{figure}
\begin{minipage}[b]{0.5\linewidth}
\begin{center}
\includegraphics[scale=0.9]{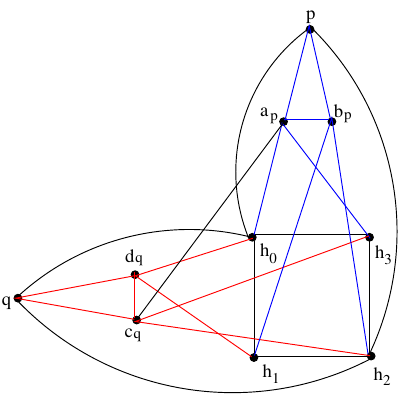}
\caption{The part of a ${\cal D}$-graph $G$ for a pair $(p,q)\in R_3$.}\label{f-III}
\end{center}
\end{minipage}
\hspace{0.5cm}
\begin{minipage}[b]{0.5\linewidth}
\begin{center}
\includegraphics[scale=0.9]{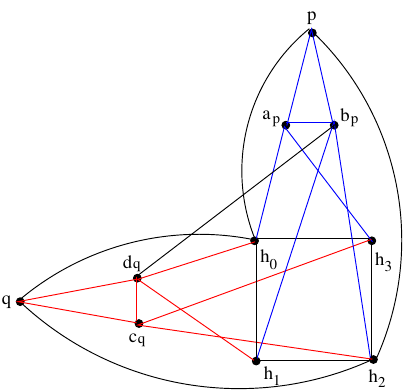}
\caption{The part of a ${\cal D}$-graph $G$ for a pair $(p,q)\in R_4$.}\label{f-IV}
\end{center}
\end{minipage}
\end{figure}

We continue the construction of $G$ by describing how we distinguish between two pairs belonging to different relations.
If $(p,q)\in R_1$, then we add the edges 
$c_qp$ and $qh_2$; see Figure~\ref{f-I}.
If $(p,q)\in R_2$, then we add the edges
$h_2p$ and $b_pq$; see Figure~\ref{f-II}.
If $(p,q)\in R_3$, then we add the edges
$h_2p$, $h_2q$ and $a_pc_q$; see Figure~\ref{f-III}.
If $(p,q)\in R_4$, then we add the edges
$h_2p$, $h_2q$ and $b_pd_q$; see Figure~\ref{f-IV}. 

We finish the construction of $G$ by adding an edge between any two vertices of type $a$, between any two vertices of type $b$, between any two vertices of type $c$, and between any two vertices of type $d$. Note that this leads to four mutually vertex-disjoint cliques in $G$; here a {\it clique} means a vertex set of a complete graph.
We call $G$ a {\it ${\cal D}$-graph} and prove the following result.

\begin{lem}\label{l-diam1}
Every ${\cal D}$-graph has diameter 2 and a dominating non-edge.
\end{lem}

\begin{pf}
Let $G$ be a ${\cal D}$-graph.
We first show that $G$ has a dominating non-edge.
Note that $h_0$ is adjacent to all vertices except to $h_2$ and 
the vertices of type 
$b$ and $c$. However, all vertices of type $b$ and $c$  are adjacent to $h_2$. 
Because $h_0$ and $h_2$ are not adjacent, this means that $h_0$ and $h_2$ form a dominating non-edge in $G$.

\begin{table}
\begin{center}
\begin{tabular}{c|c|c|c|c|c|c|c|c|c|c}
 &  $h_0$ & $h_1$ & $h_2$ & $h_3$ & $\ell$ & $r$ & $a$ & $b$ &$c$ &$d$  \\
\hline
$h_0$ & 0 & 1 & $2^{h_1}$ & 1 & 1 & 1 & 1 & $2^{h_1}$ &$2^{h_3}$ &1\\
\hline
$h_1$ & . & 0 & 1 & $2^{h_0}$ & $2^{h_0}$ & $2^{h_0}$ & $2^{h_0}$ & 1 &$2^{h_2}$ &1\\
\hline
$h_2$ & . & . & 0 & 1 & $2^{b}$ & $2^{c}$ & $2^{b}$ & 1 &1 &$2^{c}$\\
\hline
$h_3$ & . & . & . & 0 & $2^{h_0}$ & $2^{h_0}$ & 1 & $2^{a}$ &1 &$2^{c}$\\
\hline
$\ell$& . & . & . & . & $2^{h_0}$ & $2^{h_0}$ & $2^{h_0}$ & $2^{b}$ &$2^{h_2}_{c}$ &$2^{h_0}$\\
\hline
$r$   & . & . & . & . & . & $2^{h_0}$ & $2^{h_0}$ & $2^{h_2}_{b}$ &$2^{c}$ &$2^{h_0}$\\
\hline
$a$   & . & . & . & . & . & . & 1 & $2^{a}$ &$2^{h_3}$ &$2^{h_0}$\\
\hline
$b$   & . & . & . & . & . & . & . & 1 &$2^{h_2}$ &$2^{h_1}$\\
\hline
$c$   & . & . & . & . & . & . & . & . &1 &$2^{c}$\\
\hline
$d$   & . & . & . & . & . & . & . & . &. &1
\end{tabular}\\[8pt]
\caption{Determining the diameter of a ${\cal D}$-graph $G$.}\label{t-diam2}
\end{center}
\end{table}

We show that $G$ has diameter 2 in Table~\ref{t-diam2}.
In this table, $\ell$, $r$, $a$, $b$, $c$, $d$ denote vertices of corresponding type, and superscripts denote the vertex or its type that connects the two associated vertices in the case they are not adjacent already. The distances in the table must be interpreted as upper bounds.
For example, the distance between a vertex $p$ of type $\ell$ and a vertex $a_{p'}$ of type $a$ is either 1 if $p=p'$ or 2 if $p\neq p'$. 
In the latter case, they are connected by $h_0$ (and perhaps by some other vertices as well). The table denotes this as $2^{h_0}$.
In two cases the connecting vertex depends on the relation $R_i$.
Then a subscript denotes the necessary second possibility that occurs when the superscript vertex is not valid. For instance, when a vertex $p$ of type $\ell$ is not adjacent to $h_2$, then $p$ must be the first element
in a pair $(p,q)\in R_1$, and then $p$ is adjacent to $c_q$, and hence there is always an intermediate vertex
(either $h_2$ or $c_q$) to connect $p$ to an arbitrary vertex of type $c$ not necessarily equal to $c_q$. In the table this is expressed as $2^{h_2}_c$. \qed
\end{pf}

Recall that Feder and Hell~\cite{FH98} showed that $H$-{\sc Retraction} is
\NP-complete. Ito et al.~\cite{IKPT} observed that $H$-{\sc Retraction} stays
\NP-complete on graphs of diameter 3. For our purposes, we need the following theorem. Note that Lemma~\ref{l-diam1} and Theorem~\ref{t-ret} together imply that $H$-{\sc Retraction} is \NP-complete for graphs of diameter 2 that have a dominating non-edge.

\begin{thm}\label{t-ret}
The $H$-{\sc Retraction} problem is \NP-complete even for ${\cal D}$-graphs.
\end{thm}

\begin{pf}
We recall that $H$-{\sc Retraction} is in \NP, because we can guess a partition of the vertex set of the input graph $G$ into four (non-empty) sets and verify in polynomial time if this partition corresponds to a retraction from $G$ to $H$.

To show \NP-hardness, we reduce from the ${\cal D}$-{\sc Homomorphism} problem.
From an instance 
${\cal A}=(A; R_1,\ldots, R_4)$ 
of ${\cal D}$-{\sc Homomorphism}
we construct a ${\cal D}$-graph $G$.
We claim that ${\cal A}$ allows a homomorphism to ${\cal D}$ if and only if 
$G$ retracts to $H$.  

First suppose that ${\cal A}$ allows a homomorphism $f$ to ${\cal D}$.
We construct a mapping~$g$ from $V_G$ to $V_H$ as follows.
For each $a\in A$ we let $g(a)=h_i$ if $f(a)=i$, and for $i=0,\ldots, 3$ we let $g(h_i)=h_i$. 
Because $f$ is a homomorphism from ${\cal A}$ to ${\cal D}$, this leads to Tables~\ref{t-s1}--\ref{t-s4}, which explain where $a_p$, $b_p$, $c_q$ and $d_q$ map under $g$, according to where $p$ and $q$ map. From these, we conclude that $g$ is a retraction from $G$ to $H$. In particular, we note that the edges $c_qp,b_pq,a_pc_q$, and $b_pd_q$ each map to an edge or self-loop in $H$ when $(p,q)$ belongs to $R_1,\ldots,R_4$, respectively.

To prove the reverse implication, suppose that $G$ allows a retraction $g$ to $H$. We construct a mapping $f:A\to \{0,1,2,3\}$ by defining,
for each $a\in A$, $f(a)=i$  if $g(a)=h_i$.  
We claim that $f$ is a homomorphism from ${\cal A}$
to ${\cal D}$. In order to see this, we first note that $g$ maps all $a$-type vertices to $\{h_0,h_3\}$, all $b$-type vertices to $\{h_1,h_2\}$, all $c$-type vertices to $\{h_2,h_3\}$ and all $d$-type vertices to $\{h_0,h_1\}$.
We now show that $(p,q)\in R_i$ implies that $(f(p),f(q))\in S_i$ for $i=1,\ldots, 4$.

\begin{table}
\begin{minipage}[b]{0.5\linewidth}
\begin{center}
\begin{tabular}{c|c|c|c|c|c}
$p$ & $q$ & $a_p$ & $b_p$ & $c_q$ & $d_q$\\
\hline
$h_0$ & $h_3$ & $h_0$ &$h_1$ &$h_3$ &$h_0$\\
\hline
$h_1$ & $h_1$ & $h_0$ &$h_1$ &$h_2$ &$h_1$\\
\hline
$h_3$ & $h_1$ & $h_3$ &$h_2$ &$h_2$ &$h_1$\\
\hline
$h_3$ & $h_3$ & $h_3$ &$h_2$ &$h_3$ &$h_0$
\end{tabular}\\[8pt]
\caption{$g$-values when $(p,q)\in R_1$.}\label{t-s1}
\end{center}
\end{minipage}
\begin{minipage}[b]{0.5\linewidth}
\begin{center}
\begin{tabular}{c|c|c|c|c|c}
$p$ & $q$ & $a_p$ & $b_p$ & $c_q$ & $d_q$\\
\hline
$h_1$ & $h_0$ & $h_0$ &$h_1$ &$h_3$ &$h_0$\\
\hline
$h_1$ & $h_1$ & $h_0$ &$h_1$ &$h_2$ &$h_1$\\
\hline
$h_3$ & $h_1$ & $h_3$ &$h_2$ &$h_2$ &$h_1$\\
\hline
$h_3$ & $h_3$ & $h_3$ &$h_2$ &$h_3$ &$h_0$
\end{tabular}\\[8pt]
\caption{$g$-values when $(p,q)\in R_2$.}\label{t-s2}
\end{center}
\end{minipage}
\end{table}

\begin{table}
\begin{minipage}[b]{0.5\linewidth}
\begin{center}
\begin{tabular}{c|c|c|c|c|c}
$p$ & $q$ & $a_p$ & $b_p$ & $c_q$ & $d_q$\\
\hline
$h_1$ & $h_3$ & $h_0$ &$h_1$ &$h_3$ &$h_0$\\
\hline
$h_3$ & $h_1$ & $h_3$ &$h_2$ &$h_2$ &$h_1$\\
\hline
$h_3$ & $h_3$ & $h_3$ &$h_2$ &$h_3$ &$h_0$
\end{tabular}\\[8pt]
\caption{$g$-values when $(p,q)\in R_3$.}\label{t-s3}
\end{center}
\end{minipage}
\begin{minipage}[b]{0.5\linewidth}
\begin{center}
\begin{tabular}{c|c|c|c|c|c}
$p$ & $q$ & $a_p$ & $b_p$ & $c_q$ & $d_q$\\
\hline
$h_1$ &$h_1$ &$h_0$ &$h_1$ &$h_2$ &$h_1$  \\
\hline
$h_1$ & $h_3$ & $h_0$ &$h_1$ &$h_3$ &$h_0$\\
\hline
$h_3$ & $h_1$ & $h_3$ &$h_2$ &$h_2$ &$h_1$
\end{tabular}\\[8pt]
\caption{$g$-values when $(p,q)\in R_4$.}\label{t-s4}
\end{center}
\end{minipage}
\end{table}

Suppose that $(p,q)\in R_1$. Because $p$ is adjacent to $h_0$, we find that $g(p)\in \{h_0,h_1,h_3\}$. Because $q$ is adjacent to $h_0$ and $h_2$, we find that $g(q)\in \{h_1,h_3\}$. If $g(p)=h_0$, then $g$ maps $c_q$ to
$h_3$, and consequently $g(q)=h_3$.
If $g(p)=h_1$, then $g$ maps $c_q$ to $h_2$, and consequently, $d_q$ to $h_1$, implying that $g(q)=h_1$.
Hence, we find that
$(f(p),f(q))\in \{(0,3),(1,1),(3,1),(3,3)\}=S_1$, as desired.

Suppose that $(p,q)\in R_2$. 
Because $p$ is adjacent to $h_0$ and $h_2$, we find that $g(p)\in \{h_1,h_3\}$. Because $q$ is adjacent to $h_0$, we find that $g(q)\in \{h_0,h_1,h_3\}$.
If $g(q)=h_0$, then $g$ maps $b_p$ to
$h_1$, and consequently, $g(p)=h_1$.
If $g(q)=h_3$, then $g$ maps $b_p$ to $h_2$, and consequently, $a_p$ to $h_3$, implying that $g(p)=h_3$.
Hence, we find that
$(f(p),f(q))\in \{(1,0),(1,1),(3,1),(3,3)\}=S_2$, as desired.

Suppose that $(p,q)\in R_3$.
Because both $p$ and $q$ are adjacent to both $h_0$ and $h_2$, we find 
that $g(p)\in \{h_1,h_3\}$ and $g(q)\in \{h_1,h_3\}$.
If $g(p)=h_1$, then $g$ maps $a_p$ to $h_0$, and consequently, $c_q$ to $h_3$, implying that $g(q)=h_3$.
Hence, we find that
$(f(p),f(q))\in \{(1,3),(3,1),(3,3)\}=S_3$, as desired.

Suppose that $(p,q)\in R_4$.
Because both $p$ and $q$ are adjacent to both $h_0$ and $h_2$, we find 
that $g(p)\in \{h_1,h_3\}$ and $g(q)\in \{h_1,h_3\}$.
If $g(q)=h_3$, then $g$ maps $d_q$ to $h_0$, and consequently, $b_p$ to $h_1$, implying that $g(p)=h_1$.
Hence, we find that
$(f(p),f(q))\in \{(1,1),(1,3),(3,1)\}=S_4$, as desired.
This completes the proof of Lemma~\ref{t-ret}.\qed
\end{pf}

\section{Surjective Homomorphisms}\label{s-surjective}
 
\begin{figure}
\begin{center}
\includegraphics[scale=0.5]{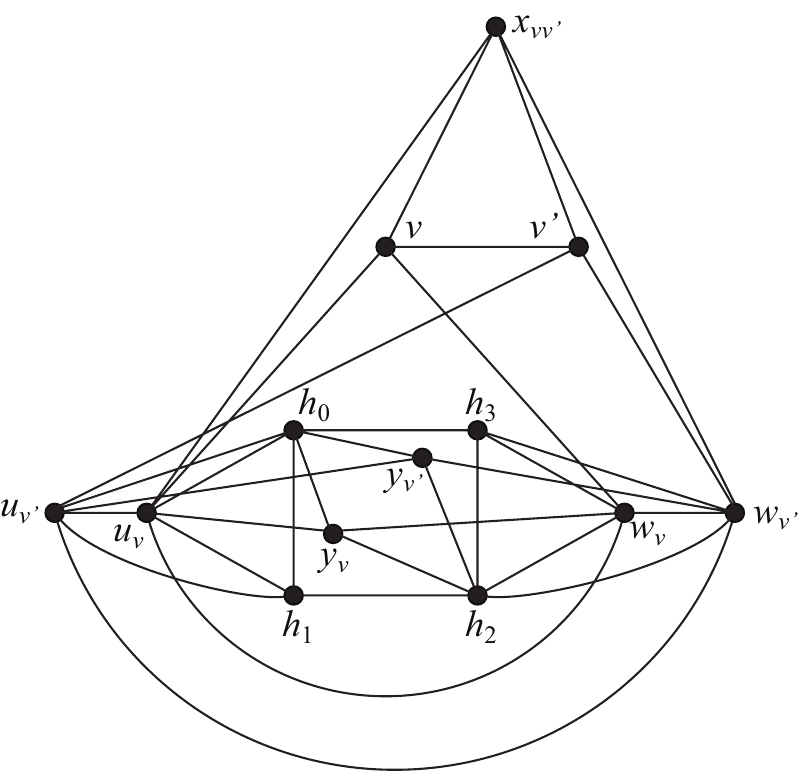}
\caption{The part of $G'$ that corresponds to edge $vv'\in E_G\setminus E_H$ as displayed in~\cite{Vi02}.}\label{f-compactor}
\end{center}
\end{figure}

Vikas~\cite{Vi02} constructed the following graph from a graph
$G=(V,E)$ that contains $H$ as an induced subgraph. 
For each vertex $v\in V_G\backslash V_H$ we add three new vertices 
$u_v,w_v,y_v$ with edges $h_0u_v, h_0y_v, h_1u_v$, $h_2w_v, h_2y_v, h_3w_v, u_vv, u_vw_v, u_vy_v$, $vw_v, w_vy_v$. 
We say that a vertex $u_v$, $w_v$ and $y_v$ has {\it type} $u$, $w$, or $y$, respectively.
We also add all edges between any two vertices $u_v,u_{v'}$ and between any two vertices $w_v,w_{v'}$ with
$v\neq v'$. For each edge $vv'$ in $E_G\backslash E_H$ we choose an arbitrary  orientation, say
from $v$ to $v'$, and then add a new vertex $x_{vv'}$ with edges $vx_{vv'}, v'x_{vv'}, u_vx_{vv'},w_{v'}x_{vv'}$. We say that this new vertex has {\it type} $x$.
The new graph $G'$ obtained from $G$ is called an $H$-{\it compactor} of $G$.
See Figure~\ref{f-compactor} for an example. This figure does not depict any self-loops, although
formally $G'$ must have at least four self-loops, because $G'$ contains $G$, and consequently, $H$ as an induced subgraph. 
However, for the same reason as for the $H$-{\sc Retraction} problem,  this is irrelevant for the {\sc Surjective $H$-Homomorphism} problem, and we may assume that $G$ and $G'$ are irreflexive.

Vikas~\cite{Vi02} showed that
a graph $G$ retracts to $H$ if and only if an (arbitrary)
$H$-compactor $G'$ of $G$ retracts to $H$ if and only if $G'$ compacts to $H$.
Recall that an $H$-compactor is of diameter 3 as observed by Ito et al.~\cite{IKPT}. Our aim is to reduce the diameter in such a graph to 2. 
This forces us to make a number of modifications. Firstly,
we must remove a number of vertices of type $x$.
Secondly, we can no longer choose the orientations regarding the remaining vertices of type $x$ arbitrarily.
Thirdly, we must connect the remaining $x$-type vertices to $H$ via edges.
We explain these modifications in detail below. 

Let $G$ be a ${\cal D}$-graph. For all vertices in $G$ we create vertices of type $u,v,w,y$ with incident edges as in the definition of a compactor. We
then perform the following three steps.

\medskip
\noindent
{\bf 1. Not creating all the vertices of type ${\mathbf x}$.}\\
We do not create $x$-type vertices for the following edges in $G$:
edges between two $a$-type vertices,
edges between two $b$-type vertices, edges between two $c$-type vertices, and edges between two $d$-type vertices. 
We create $x$-type vertices for all the other edges in $E_G\setminus E_H$ as explained in Step 2. 

\medskip
\noindent
{\bf 2. Choosing the ``right'' orientation of the other edges of $\mathbf{G-H}$.}\\
For $(p,q)\in R_i$ and $1\leq i\leq 4$, 
we choose $x$-type vertices $x_{a_pp}$, $x_{pb_p}$, $x_{a_pb_p}$,
$x_{qc_q}$, $x_{qd_q}$, and $x_{d_qc_q}$.
In addition we create the following $x$-type vertices.
For $(p,q)\in R_1$ we choose $x_{pc_q}$. 
For $(p,q)\in R_2$ we choose $x_{qb_p}$.
For $(p,q)\in R_3$ we choose $x_{a_pc_q}$.
For $(p,q)\in R_4$ we choose $x_{d_qb_p}$.
Note that in this way we have indeed created $x$-type vertices for all the other edges of
$E_G\setminus E_H$.

\medskip
\noindent
{\bf 3. Connecting the created ${\mathbf x}$-type vertices to ${\mathbf H}$.}\\
We add an edge between $h_0$ and every vertex of type $x$ that we created in Step~2. We also add an edge between $h_2$ and every such vertex.

\medskip
\noindent
We call the resulting graph a {\it semi-compactor} of $G$ and prove two
essential
lemmas.

\begin{lem}\label{l-diam2}
Let $G$ be a ${\cal D}$-graph.
Every semi-compactor of $G$ has diameter 2 and a dominating non-edge.
\end{lem}

\begin{pf}
Let $G''$ be a semi-compactor of a ${\cal D}$-graph $G$.
We first show that $G''$ has a dominating non-edge.
We note that $h_0$ is adjacent to all vertices except to $h_2$ and 
the vertices of type $b,c$, and $w$. However, all vertices of type $b,c$, and $w$ are adjacent to $h_2$. 
Because $h_0$ and $h_2$ are not adjacent, this means that $h_0$ and $h_2$ form a dominating non-edge in $G''$.

We show that $G''$ has diameter 2 in Table~\ref{t-diam3}.
In this table, $v$ denotes a vertex of $V_G$, and 
$u$, $w$, $y$, $x$ denote vertices of the corresponding type.
For reasons of clarity we explain the first row of Table~\ref{t-diam3} below;
superscripts for the other rows are used in the same way as in Table~\ref{t-diam2}.

\begin{table}
\begin{center}
\begin{tabular}{c|c|c|c|c|c}
 &  $v$ & $u$ & $w$ & $y$ & $x$ \\
\hline
$v$ & 2 & $2$ & $2$ & $2$ & $2$\\
\hline
$u$ & . & 1 & $2^w$ & $2^{h_0}$ & $2^{h_0}$\\
\hline
$w$ & . & . & 1 & $2^{h_2}$ & $2^{h_2}$\\
\hline
$y$ & . & . & . & $2^{h_0}$ & $2^{h_0}$\\
\hline
$x$& . & . & . & . & $2^{h_0}$ 
\end{tabular}\\[8pt]
\caption{Determining the diameter of a semi-compactor $G''$.}\label{t-diam3}
\end{center}
\end{table}

For the first position of row 1 we use Table~\ref{t-diam2}
to determine an upper bound for the distance between two vertices in $G''$; hence, there is no superscript for this position.
For the second position of row 1 we use the fact that every vertex $v\in V_G\setminus V_H$ is adjacent to $u_v$ and that $u_v$ is adjacent to every
other vertex of type $u$. Furthermore, if $v=h_0$ or $v=h_1$ then $v$ is adjacent
to every vertex of type $u$, and if $v=h_2$ or $v=h_3$ then $v$ is of distance two from every vertex of type $u$ by using $h_0$ or $h_1$ as an intermediate vertex, respectively. The third position of row 1 can be explained by similar arguments. The fourth and fifth positions follow from the already deduced property of $G''$ that
$h_0$ and $h_2$ form a dominating non-edge combined with the property that
every vertex of type $x$ and $y$ is adjacent to both $h_0$ and $h_2$.
This completes the proof of Lemma~\ref{l-diam2}.\qed
\end{pf}

\begin{lem}\label{l-equi}
Let $G''$ be a semi-compactor of a ${\cal D}$-graph $G$.
Then the following statements are equivalent:
\begin{itemize}
\item [(i)] $G$ retracts to $H$;
\item [(ii)] $G''$ retracts to $H$;
\item [(iii)] $G''$ compacts to $H$;
\item [(iv)] $G''$ has a vertex-surjective homomorphism to $H$.
\end{itemize}
\end{lem}

\begin{pf}
We show the following implications: $(i)\Rightarrow (ii)$, $(ii)\Rightarrow (i)$, $(ii)\Rightarrow (iii)$, $(iii)\Rightarrow (ii)$, $(iii)\Rightarrow (iv)$,
and $(iv)\Rightarrow (iii)$.

\medskip
\noindent
``$(i)\Rightarrow (ii)$'' Let $f$ be a retraction from $G$ to $H$. We show how to extend 
$f$ to a retraction from $G''$ to $H$. We observe that every vertex of type $u$ can only be mapped to $h_0$ or $h_1$, because such a vertex is adjacent to $h_0$ and $h_1$.
We also observe that every vertex of type $w$ can only be mapped to $h_2$ or $h_3$,
because such a vertex is adjacent to $h_2$ and $h_3$.
This implies the following.
Let $v\in V_G\setminus V_H$. 
If $f(v)=h_0$ or $f(v)=h_1$, then $w_v$ must be mapped to $h_3$ or $h_2$, respectively. Consequently,
$u_v$ must be mapped to $h_0$ or $h_1$, respectively, due to the edge $u_vw_v$.
If $f(v)=h_2$ or $f(v)=h_3$, 
then $u_v$ must be mapped to $h_1$ or $h_0$, respectively. 
Consequently, $w_v$ must be mapped to $h_2$ or $h_3$, respectively, due to the edge $u_vw_v$.
Hence, $f(v)$ fixes the mapping of the vertices $u_v$ and $w_v$. Moreover, we showed that 
either $u_v$ is mapped to $h_1$ or $w_v$ is mapped to $h_3$. 
Note that both vertices are adjacent to $y_v$.
Then, because $y_v$ can only be mapped to $h_1$ or $h_3$ due to the edges $h_0y_v$ and 
$h_2y_v$, the mapping of $y_v$ is fixed as well; if $u_v$ is mapped to $h_1$ then $y_v$ is mapped to $h_1$, and if $w_v$ is mapped to $h_3$ then
$y_v$ is mapped to $h_3$. 

What is left to do is to verify whether we can map the vertices of type $x$.
For this purpose we refer to Table~\ref{t-fixed}, where $v,v'$ denote two adjacent 
vertices of $V_G\setminus V_H$. Every possible combination of $f(v)$ and $f(v')$ corresponds to a row in this table. As we have just shown, this fixes the image of 
the vertices $u_v$, $u_{v'}$, $w_v$, $w_{v'}$, $y_{v'}$ and $y_v$. 
For $x_{vv'}$ we use its adjacencies to $v$, $v'$, $u_v$ and $w_{v'}$ to determine
potential images. For some cases, this number of potential images is not one but two.
This is shown in the last column of Table~\ref{t-fixed}; here we did not take into account that every $x_{vv'}$ is adjacent to $h_0$ and $h_2$ in our construction.
Because of these adjacencies, every $x_{vv'}$ can only be mapped to $h_1$ or $h_3$. 
In the majority of the 12 rows in Table~\ref{t-fixed} we have this choice;
the exceptions are row 4 and row 9. In rows 4 and 9, we find that $x_{vv'}$ can only be mapped to one image, which is $h_0$ or $h_2$, respectively.
We will show that neither row can occur.

\begin{table}
\begin{center}
\begin{tabular}{c|c|c|c|c|c|c|c|c}
   $v$ & $v'$ & $u_v$ & $u_{v'}$ & $w_v$ & $w_{v'}$ & $y_v$ & $y_{v'}$ &$x_{vv'}$\\
   \hline
   $h_0$ &$h_0$ &$h_0$ &$h_0$ &$h_3$ &$h_3$ &$h_3$ &$h_3$ &$h_0/h_3$\\
   \hline
   $h_0$ &$h_1$ &$h_0$ &$h_1$ &$h_3$ &$h_2$ &$h_3$ &$h_1$ &$h_1$\\
   \hline
   $h_0$ &$h_3$ &$h_0$ &$h_0$ &$h_3$ &$h_3$ &$h_3$ &$h_3$ &$h_0/h_3$\\
   \hline
   $h_1$ &$h_0$ &$h_1$ &$h_0$ &$h_2$ &$h_3$ &$h_1$ &$h_3$ &$h_0$\\
   \hline
   $h_1$ &$h_1$ &$h_1$ &$h_1$ &$h_2$ &$h_2$ &$h_1$ &$h_1$ &$h_1/h_2$\\
   \hline  
    $h_1$ &$h_2$ &$h_1$ &$h_1$ &$h_2$ &$h_2$ &$h_1$ &$h_1$ &$h_1/h_2$\\
   \hline
    $h_2$ &$h_1$ &$h_1$ &$h_1$ &$h_2$ &$h_2$ &$h_1$ &$h_1$ &$h_1/h_2$\\
   \hline
    $h_2$ &$h_2$ &$h_1$ &$h_1$ &$h_2$ &$h_2$ &$h_1$ &$h_1$ &$h_1/h_2$\\
   \hline
    $h_2$ &$h_3$ &$h_1$ &$h_0$ &$h_2$ &$h_3$ &$h_1$ &$h_3$ &$h_2$\\
   \hline
    $h_3$ &$h_0$ &$h_0$ &$h_0$ &$h_3$ &$h_3$ &$h_3$ &$h_3$ &$h_0/h_3$\\
   \hline
    $h_3$ &$h_2$ &$h_0$ &$h_1$ &$h_3$ &$h_2$ &$h_3$ &$h_1$ &$h_3$\\
   \hline
   $h_3$ &$h_3$ &$h_0$ &$h_0$ &$h_3$ &$h_3$ &$h_3$ &$h_3$ &$h_0/h_3$
\end{tabular}\\[8pt]
\caption{Determining a retraction from $G''$ to $H$.}\label{t-fixed}
\end{center}
\end{table}

By Steps 1-2 of the definition of a semi-compactor, we have that $(v,v')$ belongs to
$$\{(a_p,p), (p,b_p), (a_p,b_p), (q,c_q), (q,d_q), (d_q,c_q), (p,c_q), (q,b_p),(a_p,c_q),(d_q,b_p)\}.$$
We first show that row 4 cannot occur. In order to obtain a contradiction, suppose that row 4 does occur, i.e., that $f(v)=h_1$ and $f(v')=h_0$ for some $v,v'\in V_G\setminus V_H$.
Due to their adjacencies with vertices of $H$, every vertex of type $a$ is mapped to
$h_0$ or $h_3$, every vertex of type $b$ to $h_1$ or $h_2$, every vertex of type $c$ to 
$h_2$ or $h_3$ and every vertex of type $d$ to $h_0$ or $h_1$.
This means that $v$ can only be $p,q,b_p$, or $d_q$, whereas $v'$ can only be 
$p$, $q$, $a_p$ or $d_q$. If $v=p$ then $v'\in \{b_p,c_q\}$. If
$v=q$ then $v'\in \{c_q,d_q,b_p\}$. If $v=b_p$ then $v'$ cannot be chosen.
If $v=d_q$ then $v'\in \{c_q,b_p\}$.
Hence, we find that $v=q$ and $v'=d_q$. However, then $f$ is not a retraction from 
$G$ to $H$, because $c_q$ is adjacent to $d_q,q,h_2,h_3$, and $f$ maps these vertices to $h_0,h_1,h_2,h_3$, respectively. Hence, row 4 does not occur.

We now show that row 9 cannot occur. In order to obtain a contradiction, suppose that row 9 does occur, i.e., that $f(v)=h_2$ and $f(v')=h_3$. 
As in the previous case, we deduce that every vertex of type $a$ is mapped to
$h_0$ or $h_3$, every vertex of type $b$ to $h_1$ or $h_2$, every vertex of type $c$ to 
$h_2$ or $h_3$ and every vertex of type $d$ to $h_0$ or $h_1$. Moreover, every vertex
of type $\ell$ or $r$ cannot be mapped to $h_2$, because it is adjacent to $h_0$.
Then $v$ can only be $b_p$ or $c_q$, and $v'$ can only be $p$, $q$, $a_p$ or
$c_q$. However, if $v=b_p$ or $v=c_q$ then $v'$ cannot be chosen.
Hence, row 9 cannot occur, and we conclude that $f$ can be extended to a retraction from $G''$ to $H$, as desired.

\medskip
\noindent
``$(ii)\Rightarrow (i)$'' Let $f$ be a retraction from $G''$ to $H$.
Then the restriction of $f$ to $V_{G}$ is a retraction from $G$ to $H$.
Hence, this implication is valid.

\medskip
\noindent
``$(ii)\Rightarrow (iii)$'' This implication is valid, because every retraction from $G''$ to $H$ is an edge-surjective homomorphism, so \emph{a fortiori} a compaction from $G''$ to $H$.

\medskip
\noindent
``$(iii)\Rightarrow (ii)$'' Let $f$ be a compaction from $G''$ to $H$. We will show
that $f$ is without loss of generality a retraction from $G''$ to $H$.
Our proof goes along the same lines as the proof of Lemma 2.1.2 in Vikas~\cite{Vi02}, i.e., we use the same arguments but in addition
we must examine a few more cases due to our modifications in steps 1--3; we therefore include all the proof details below.

We let $U$ consist of $h_0,h_1$ and all vertices of type $u$. Similarly, we let 
$W$ consist of $h_2,h_3$ and all vertices of type $w$.
Because $U$ forms a clique in $G$, we find that $f(U)$ is a clique in $H$.
This means that $1\leq |f(U)|\leq 2$. By the same arguments, we find that
$1\leq f(W)\leq 2$. 

We first prove that $|f(U)|=|f(W)|=2$.
In order to derive a contradiction, suppose that 
$|f(U)|\neq 2$. Then $f(U)$ has only one vertex. By symmetry, we may assume that $f$ maps every vertex of $U$ to $h_0$; otherwise we can redefine $f$.
Because every vertex of $G''$ is adjacent
to a vertex in $U$, we find that
$G''$ contains no vertex that is mapped to $h_2$ by $f$. This is not possible, because
$f$ is a compaction from $G''$ to $H$.
Hence $|f(U)|=2$, and by the same arguments, $|f(W)|=2$.
Because $U$ is a clique, we find that $f(U)\neq \{h_0,h_2\}$ and $f(U)\neq \{h_1,h_3\}$.
Hence, by symmetry, we assume that $f(U)=\{h_0,h_1\}$.

We now prove that $f(W)=\{h_2,h_3\}$. In order to obtain a contradiction, suppose that
$f(W)\neq \{h_2,h_3\}$.
Because $f$ is a compaction from $G''$ to $H$, there exists an 
edge $st$ in $G''$ with $f(s)=h_2$ and $f(t)=h_3$.
Because $f(U)$ only contains vertices mapped to $h_0$ or $h_1$, we find that
$s\notin U$ and $t\notin U$.
Because we assume that $f(W)\neq \{h_2,h_3\}$, we find that 
$st$ is not one of 
$w_vw_{v'},w_vh_2,w_vh_3,h_2h_3$. Hence, $st$ 
is one of the following edges
$$vw_v, w_vy_v, vx_{vv'}, y_vh_2, vh_2, vh_3,
vv', v'x_{vv'},w_{v'}x_{vv'},x_{vv'}h_2,$$
where $v,v'\in V_G\setminus V_H$. We must consider each of these possibilities.

If $st\in \{vw_v,w_vy_v,vx_{vv'}\}$ then $f(u_v)\in \{h_2,h_3\}$, because $u_v$ is adjacent to $v,w_v,y_v,x_{vv'}$.
However, this is not possible because $f(u_v)\in \{h_0,h_1\}$.

If $st=y_vh_2$, then $f(w_v)=h_2$ or $f(w_v)=h_3$, because $w_v$ is adjacent to 
both $y_v$ and $h_2$, and
$\{f(y_v),f(h_2)\}=\{h_2,h_3\}$.
This means that either $f(w_v)=f(y_v)$ or $f(w_v)=f(h_2)$.
If $f(w_v)=f(y_v)$, then $\{f(w_v),f(h_2)\}=\{h_2,h_3\}$.
Consequently,  $f(W)=\{h_2,h_3\}$, which we assumed is not the case.
Hence, $f(w_v)\neq f(y_v)$. Then $f$ maps the edge $w_vy_v$ to $h_2h_3$, and we return to the previous case. We can repeat the same arguments if $st=vh_2$ or $st=vh_3$.
Hence, we find that $st$ cannot be equal to those edges either.

If $st=vv'$, then by symmetry we may assume without loss of generality that
$f(v)=h_2$ and $f(v')=h_3$. Consequently, $f(u_v)=h_1$, because $u_v\in U$ is adjacent to $v$, and can only be mapped to $h_0$ or $h_1$.
By the same reasoning, $f(u_{v'})=h_0$. Because $w_v$ is adjacent to $v$ with $f(v)=h_2$
and to $u_v$ with $f(u_v)=h_1$, we find that $f(w_v)\in \{h_1,h_2\}$.
Because $w_{v'}$ is adjacent to $v'$ with $f(v')=h_3$ and to $u_{v'}$ with 
$f(u_{v'})=h_0$, we find that $f(w_{v'})\in \{h_0,h_3\}$. Recall that $f(W)\neq \{h_2,h_3\}$. Then, because $w_v$ and $w_{v'}$ are adjacent, we find that $f(w_v)=h_1$ and $f(w_{v'})=h_0$. Suppose that $x_{vv'}$ exists. Then $x_{vv'}$ is adjacent to vertices 
$v$ with $f(v)=h_2$, to $v'$ with $f(v')=h_3$, to $u_v$ with $f(u_v)=h_1$ and to
$w_{v'}$ with $f(w_{v'})=h_0$. This is not possible. Hence $x_{vv'}$ cannot exist.
This means that $v,v'$ are both of type $a$, both of type $b$, both of type $c$ or both of type $d$. If $v,v'$ are both of type $a$ or both of type $d$, then $f(h_0)\in \{h_2,h_3\}$, which is not possible because $h_0\in U$ and $f(U)=\{h_0,h_1\}$. 
If $v,v'$ are both of type $b$, we apply the same reasoning with respect to
$h_1$. Suppose that $v,v'$ are both of type $c$. Then both $v$ and $v'$ are adjacent to $h_2$. This means that $f(h_2)\in \{h_2,h_3\}$.
Then either $\{f(v),f(h_2)\}=\{h_2,h_3\}$ or $\{f(v'),f(h_2)\}=\{h_2,h_3\}$.
Hence, by considering either the edge $vh_2$ or $v'h_2$ we return to a previous case. We conclude that $st\neq vv'$.

If $st=v'x_{vv'}$ then $f(v)\in \{h_2,h_3\}$, because $v$ is adjacent to $v'$ and $x_{vv'}$. Then one of $vv'$ or $vx_{vv'}$ maps to $h_2h_3$, and we can return to a previous case. Hence, we find that $st\neq v'x_{vv'}$.

If $st=w_{v'}x_{vv'}$ then $f(v')\in \{h_2,h_3\}$, because $v'$ is adjacent to $w_{v'}$ and 
$x_{vv'}$. Then one of $v'w_{v'}$ or $v'x_{vv'}$ maps to $h_2h_3$, and we can return to a previous case. Hence, we find that $st\neq w_{v'}x_{vv'}$.

If $st=x_{vv'}h_2$ then $f(w_{v'})\in \{h_2,h_3\}$, because $w_{v'}$ is adjacent to 
$x_{vv'}$ and $h_2$. Because $f(W)\neq \{h_2,h_3\}$, we find that $f(w_{v'})=f(h_2)$.
Then $w_{v'}x_{vv'}$ is mapped to $h_2h_3$, and we return to a previous case.
Hence, we find that $st\neq x_{vv'}h_2$.

We conclude that $G''$ has no edge $st$ with $f(s)=h_2$ and $f(t)=h_3$. 
This is a contradiction; recall that $f$ is a compaction. Hence, $f(W)=\{h_2,h_3\}$.

The next step is to prove that $f(h_0)\neq f(h_1)$. In order to obtain a contradiction, suppose that $f(h_0)=f(h_1)$. By symmetry we may assume without loss of generality that $f(h_0)=f(h_1)=h_0$. Because $f(U)=\{h_0,h_1\}$, there exists a vertex $u_v$  with
$f(u_v)=h_1$. Because $w_v$ with $f(w_v)\in \{h_2,h_3\}$ is adjacent to $u_v$, we find 
that $f(w_v)=h_2$. Because $h_2$ with $f(h_2)\in \{h_2,h_3\}$ is adjacent to $h_1$ with $f(h_1)=h_0$, we find that $f(h_2)=h_3$. However, then $y_v$ is adjacent to 
$h_0$ with $f(h_0)=h_0$, to $u_v$ with $f(u_v)=h_1$, to $w_v$ with $f(w_v)=h_2$, and
to $h_2$ with $f(h_2)=h_3$. This is not possible.
Hence, we find that $f(h_0)\neq f(h_1)$. By symmetry, we may assume without loss of generality that $f(h_0)=h_0$ and $f(h_1)=h_1$.

We are left to show that $f(h_2)=h_2$ and $f(h_3)=h_3$. 
This can be seen as follows.
Because $h_2$ is adjacent to $h_1$ with $f(h_1)=h_1$, and $f(h_2)\in \{h_2,h_3\}$ we find that
$f(h_2)=h_2$. 
Because $h_3$ is adjacent to $h_0$ with $f(h_0)=h_0$, and $f(h_3)\in \{h_2,h_3\}$ we find that $f(h_3)=h_3$. Hence, we have found that $f$ is a retraction from $G''$ to $H$, as desired.

\medskip
\noindent
``$(iii)\Rightarrow (iv)$'' and ``$(iv)\Rightarrow (iii)$'' immediately follow from the equivalence between statements 3 and 6 in Proposition~\ref{p-known},
after recalling that $G''$ has diameter 2 due to Lemma~\ref{l-diam2}.
\qed
\end{pf}
We are now ready to state the main result of our paper. Its proof follows from 
Lemmas~\ref{l-diam2} and~\ref{l-equi}, in light of Theorem~\ref{t-ret};
note that all constructions may be carried out in polynomial time.

\begin{thm}\label{t-main}
The {\sc Surjective ${\mathcal C}_4$-Homomorphism} problem is \NP-complete for 
graphs of diameter 2 even if they have a dominating non-edge.
\end{thm}

\medskip
\noindent
{\bf Acknowledgments.} The authors thank Andrei Krokhin for useful comments on Section~\ref{s-algebra} and  
an anonymous reviewer for helpful comments on the presentation of our paper.

\bibliographystyle{elsart-num}

\end{document}